\documentclass[manuscript]{emulateapj}

\usepackage{natbib}
\usepackage{lineno}
\usepackage{multirow}
\usepackage{booktabs}
\usepackage{epsfig}
\usepackage{color}
\usepackage{ulem}
\usepackage{amsmath}
\usepackage{rotating}

\usepackage[colorinlistoftodos]{todonotes}

\shorttitle{Galaxies with Multiple SNe Ia}
\shortauthors{Scolnic, Smith, Massiah et al.}

\begin{document}

\begin{nolinenumbers}
\vspace*{-\headsep}\vspace*{\headheight}
\footnotesize \hfill FERMILAB-PUB-20-057-AE\\
\vspace*{-\headsep}\vspace*{\headheight}
\footnotesize \hfill DES-2019-0505
\end{nolinenumbers}

\title{Supernova Siblings: Assessing the consistency of properties of Type Ia Supernovae that share the same parent galaxies}

\def\andname{}

\author{
D.~Scolnic\altaffilmark{1},
M.~Smith\altaffilmark{2},
A.~Massiah\altaffilmark{3},
P.~Wiseman\altaffilmark{2},
D.~Brout\altaffilmark{4,5},
R.~Kessler\altaffilmark{6,7},
T.~M.~Davis\altaffilmark{8},
R.~J.~Foley\altaffilmark{9},
L.~Galbany\altaffilmark{10},
S.~R.~Hinton\altaffilmark{8},
R.~Hounsell\altaffilmark{4},
L.~Kelsey\altaffilmark{2},
C.~Lidman\altaffilmark{11},
E.~Macaulay\altaffilmark{12},
R.~Morgan\altaffilmark{13},
R.~C.~Nichol\altaffilmark{12},
A.~M\"oller\altaffilmark{14},
B.~Popovic\altaffilmark{1},
M.~Sako\altaffilmark{4},
M.~Sullivan\altaffilmark{2},
B.~P.~Thomas\altaffilmark{12},
B.~E.~Tucker\altaffilmark{11},
T.~M.~C.~Abbott\altaffilmark{15},
M.~Aguena\altaffilmark{16,17},
S.~Allam\altaffilmark{18},
J.~Annis\altaffilmark{18},
S.~Avila\altaffilmark{19},
K.~Bechtol\altaffilmark{20,13},
E.~Bertin\altaffilmark{21,22},
D.~Brooks\altaffilmark{23},
D.~L.~Burke\altaffilmark{24,25},
A.~Carnero~Rosell\altaffilmark{26},
D.~Carollo\altaffilmark{27},
M.~Carrasco~Kind\altaffilmark{28,29},
J.~Carretero\altaffilmark{30},
M.~Costanzi\altaffilmark{31,32},
L.~N.~da Costa\altaffilmark{17,33},
J.~De~Vicente\altaffilmark{26},
S.~Desai\altaffilmark{34},
H.~T.~Diehl\altaffilmark{18},
P.~Doel\altaffilmark{23},
A.~Drlica-Wagner\altaffilmark{6,18,7},
K.~Eckert\altaffilmark{4},
T.~F.~Eifler\altaffilmark{35,36},
S.~Everett\altaffilmark{9},
B.~Flaugher\altaffilmark{18},
P.~Fosalba\altaffilmark{37,38},
J.~Frieman\altaffilmark{18,7},
J.~Garc\'ia-Bellido\altaffilmark{19},
E.~Gaztanaga\altaffilmark{37,38},
D.~W.~Gerdes\altaffilmark{39,40},
K.~Glazebrook\altaffilmark{41},
D.~Gruen\altaffilmark{42,24,25},
R.~A.~Gruendl\altaffilmark{28,29},
J.~Gschwend\altaffilmark{17,33},
G.~Gutierrez\altaffilmark{18},
W.~G.~Hartley\altaffilmark{23,43},
D.~L.~Hollowood\altaffilmark{9},
K.~Honscheid\altaffilmark{44,45},
D.~J.~James\altaffilmark{46},
K.~Kuehn\altaffilmark{47,48},
N.~Kuropatkin\altaffilmark{18},
G.~F.~Lewis\altaffilmark{49},
T.~S.~Li\altaffilmark{50,51},
M.~Lima\altaffilmark{16,17},
M.~A.~G.~Maia\altaffilmark{17,33},
J.~L.~Marshall\altaffilmark{52},
F.~Menanteau\altaffilmark{28,29},
R.~Miquel\altaffilmark{53,30},
A.~Palmese\altaffilmark{18,7},
F.~Paz-Chinch\'{o}n\altaffilmark{28,29},
A.~A.~Plazas\altaffilmark{50},
M.~Pursiainen\altaffilmark{2},
E.~Sanchez\altaffilmark{26},
V.~Scarpine\altaffilmark{18},
M.~Schubnell\altaffilmark{40},
S.~Serrano\altaffilmark{37,38},
I.~Sevilla-Noarbe\altaffilmark{26},
N.~E.~Sommer\altaffilmark{11},
E.~Suchyta\altaffilmark{54},
M.~E.~C.~Swanson\altaffilmark{29},
G.~Tarle\altaffilmark{40},
T.~N.~Varga\altaffilmark{55,56},
A.~R.~Walker\altaffilmark{15},
and R.~Wilkinson\altaffilmark{57}
\\ \vspace{0.2cm} (DES Collaboration) \\
}

\affil{$^{1}$ Department of Physics, Duke University Durham, NC 27708, USA}
\email{daniel.scolnic@duke.edu}
\affil{$^{2}$ School of Physics and Astronomy, University of Southampton,  Southampton, SO17 1BJ, UK}
\affil{$^{3}$ University of Connecticut, Storrs, CT 06269, USA}
\affil{$^{4}$ Department of Physics and Astronomy, University of Pennsylvania, Philadelphia, PA 19104, USA}
\affil{$^{5}$ NASA Einstein Fellow}
\affil{$^{6}$ Department of Astronomy and Astrophysics, University of Chicago, Chicago, IL 60637, USA}
\affil{$^{7}$ Kavli Institute for Cosmological Physics, University of Chicago, Chicago, IL 60637, USA}
\affil{$^{8}$ School of Mathematics and Physics, University of Queensland,  Brisbane, QLD 4072, Australia}
\affil{$^{9}$ Santa Cruz Institute for Particle Physics, Santa Cruz, CA 95064, USA}
\affil{$^{10}$ PITT PACC, Department of Physics and Astronomy, University of Pittsburgh, Pittsburgh, PA 15260, USA}
\affil{$^{11}$ The Research School of Astronomy and Astrophysics, Australian National University, ACT 2601, Australia}
\affil{$^{12}$ Institute of Cosmology and Gravitation, University of Portsmouth, Portsmouth, PO1 3FX, UK}
\affil{$^{13}$ Physics Department, 2320 Chamberlin Hall, University of Wisconsin-Madison, 1150 University Avenue Madison, WI  53706-1390}
\affil{$^{14}$ Universit'e Clermont Auvergne, CNRS/IN2P3, LPC, F-63000 Clermont-Ferrand, France}
\affil{$^{15}$ Cerro Tololo Inter-American Observatory, National Optical Astronomy Observatory, Casilla 603, La Serena, Chile}
\affil{$^{16}$ Departamento de F\'isica Matem\'atica, Instituto de F\'isica, Universidade de S\~ao Paulo, CP 66318, S\~ao Paulo, SP, 05314-970, Brazil}
\affil{$^{17}$ Laborat\'orio Interinstitucional de e-Astronomia - LIneA, Rua Gal. Jos\'e Cristino 77, Rio de Janeiro, RJ - 20921-400, Brazil}
\affil{$^{18}$ Fermi National Accelerator Laboratory, P. O. Box 500, Batavia, IL 60510, USA}
\affil{$^{19}$ Instituto de Fisica Teorica UAM/CSIC, Universidad Autonoma de Madrid, 28049 Madrid, Spain}
\affil{$^{20}$ LSST, 933 North Cherry Avenue, Tucson, AZ 85721, USA}
\affil{$^{21}$ CNRS, UMR 7095, Institut d'Astrophysique de Paris, F-75014, Paris, France}
\affil{$^{22}$ Sorbonne Universit\'es, UPMC Univ Paris 06, UMR 7095, Institut d'Astrophysique de Paris, F-75014, Paris, France}
\affil{$^{23}$ Department of Physics \& Astronomy, University College London, Gower Street, London, WC1E 6BT, UK}
\affil{$^{24}$ Kavli Institute for Particle Astrophysics \& Cosmology, P. O. Box 2450, Stanford University, Stanford, CA 94305, USA}
\affil{$^{25}$ SLAC National Accelerator Laboratory, Menlo Park, CA 94025, USA}
\affil{$^{26}$ Centro de Investigaciones Energ\'eticas, Medioambientales y Tecnol\'ogicas (CIEMAT), Madrid, Spain}
\affil{$^{27}$ INAF, Astrophysical Observatory of Turin, I-10025 Pino Torinese, Italy}
\affil{$^{28}$ Department of Astronomy, University of Illinois at Urbana-Champaign, 1002 W. Green Street, Urbana, IL 61801, USA}
\affil{$^{29}$ National Center for Supercomputing Applications, 1205 West Clark St., Urbana, IL 61801, USA}
\affil{$^{30}$ Institut de F\'{\i}sica d'Altes Energies (IFAE), The Barcelona Institute of Science and Technology, Campus UAB, 08193 Bellaterra (Barcelona) Spain}
\affil{$^{31}$ INAF-Osservatorio Astronomico di Trieste, via G. B. Tiepolo 11, I-34143 Trieste, Italy}
\affil{$^{32}$ Institute for Fundamental Physics of the Universe, Via Beirut 2, 34014 Trieste, Italy}
\affil{$^{33}$ Observat\'orio Nacional, Rua Gal. Jos\'e Cristino 77, Rio de Janeiro, RJ - 20921-400, Brazil}
\affil{$^{34}$ Department of Physics, IIT Hyderabad, Kandi, Telangana 502285, India}
\affil{$^{35}$ Department of Astronomy/Steward Observatory, University of Arizona, 933 North Cherry Avenue, Tucson, AZ 85721-0065, USA}
\affil{$^{36}$ Jet Propulsion Laboratory, California Institute of Technology, 4800 Oak Grove Dr., Pasadena, CA 91109, USA}
\affil{$^{37}$ Institut d'Estudis Espacials de Catalunya (IEEC), 08034 Barcelona, Spain}
\affil{$^{38}$ Institute of Space Sciences (ICE, CSIC),  Campus UAB, Carrer de Can Magrans, s/n,  08193 Barcelona, Spain}
\affil{$^{39}$ Department of Astronomy, University of Michigan, Ann Arbor, MI 48109, USA}
\affil{$^{40}$ Department of Physics, University of Michigan, Ann Arbor, MI 48109, USA}
\affil{$^{41}$ Centre for Astrophysics \& Supercomputing, Swinburne University of Technology, Victoria 3122, Australia}
\affil{$^{42}$ Department of Physics, Stanford University, 382 Via Pueblo Mall, Stanford, CA 94305, USA}
\affil{$^{43}$ Department of Physics, ETH Zurich, Wolfgang-Pauli-Strasse 16, CH-8093 Zurich, Switzerland}
\affil{$^{44}$ Center for Cosmology and Astro-Particle Physics, The Ohio State University, Columbus, OH 43210, USA}
\affil{$^{45}$ Department of Physics, The Ohio State University, Columbus, OH 43210, USA}
\affil{$^{46}$ Center for Astrophysics $\vert$ Harvard \& Smithsonian, 60 Garden Street, Cambridge, MA 02138, USA}
\affil{$^{47}$ Australian Astronomical Optics, Macquarie University, North Ryde, NSW 2113, Australia}
\affil{$^{48}$ Lowell Observatory, 1400 Mars Hill Rd, Flagstaff, AZ 86001, USA}
\affil{$^{49}$ Sydney Institute for Astronomy, School of Physics, A28, The University of Sydney, NSW 2006, Australia}
\affil{$^{50}$ Department of Astrophysical Sciences, Princeton University, Peyton Hall, Princeton, NJ 08544, USA}
\affil{$^{51}$ Observatories of the Carnegie Institution for Science, 813 Santa Barbara St., Pasadena, CA 91101, USA}
\affil{$^{52}$ George P. and Cynthia Woods Mitchell Institute for Fundamental Physics and Astronomy, and Department of Physics and Astronomy, Texas A\&M University, College Station, TX 77843,  USA}
\affil{$^{53}$ Instituci\'o Catalana de Recerca i Estudis Avan\c{c}ats, E-08010 Barcelona, Spain}
\affil{$^{54}$ Computer Science and Mathematics Division, Oak Ridge National Laboratory, Oak Ridge, TN 37831}
\affil{$^{55}$ Max Planck Institute for Extraterrestrial Physics, Giessenbachstrasse, 85748 Garching, Germany}
\affil{$^{56}$ Universit\"ats-Sternwarte, Fakult\"at f\"ur Physik, Ludwig-Maximilians Universit\"at M\"unchen, Scheinerstr. 1, 81679 M\"unchen, Germany}
\affil{$^{57}$ Department of Physics and Astronomy, Pevensey Building, University of Sussex, Brighton, BN1 9QH, UK}

\submitted{Submitted to The Astrophysical Journal Letters}

\begin{abstract}
While many studies have shown a correlation between properties of the light curves of Type Ia SN (SNe Ia) and properties of their host galaxies, it remains unclear what is driving these correlations.  We introduce a new direct method to study these correlations by analyzing `parent' galaxies that host multiple SNe Ia `siblings'.  Here, we search the Dark Energy Survey SN sample, one of the largest samples of discovered SNe, and find 8 galaxies that hosted two likely Type Ia SNe.  Comparing the light-curve properties of these SNe and recovered distances from the light curves, we find no better agreement between properties of SNe in the same galaxy as any random pair of galaxies, with the exception of the SN light-curve stretch.  We show at $2.8\sigma$ significance that at least 1/2 of the intrinsic scatter of SNe Ia distance modulus residuals is not from common host properties.  We also discuss the robustness with which we could make this evaluation with LSST, which will find $100\times$ more pairs of galaxies, and  pave a new line of study on the consistency of Type Ia supernovae in the same parent galaxies.  Finally, we argue that it is unlikely some of these SNe are actually single, lensed SN with multiple images.
\end{abstract}

\section{Introduction}
Analyses of increasingly large samples of supernovae have revealed correlations between properties of Type Ia supernovae (SNe Ia) light curves and properties of their host galaxies.  The light-curve widths have been shown to correlate with host-galaxy morphology (e.g., \citealp{Hamuy1996}), mass (e.g., \citealp{Howell2009}) and star-formation rate (e.g., \citealp{Sullivan2006,Smith12}).  The light-curve color has been shown to correlate weakly with host-galaxy metallicity (e.g, \citealp{Childress2013}) and host-galaxy mass (e.g., \citealp{Brout18-SYS}).  After standardizing SN brightness using a light-curve model like SALT2 \citep{Guy2007} or MLCS2k2 \citep{Jha2007}, multiple analyses have also shown a correlation between host-galaxy mass and the distance modulus residuals of the SNe Ia relative to the best-fit cosmology (e.g., \citealp{Kelly2010,Lampeitl2010,Sullivan2010}.  Similar correlations between host properties and SN distance modulus residuals have been shown using star formation (e.g, \citealp{DAndrea2011}) and metallicity (e.g., \citealp{Hayden2013}). {As galaxy demographics are known to evolve with redshift \citep{Childress2013}, analyses that measure the dark energy equation-of-state $w$ with SNe Ia must account for the relationship between host-galaxy properties and SN light-curve properties in order to reduce systematic uncertainties in the cosmological measurement.}
\color{black}{}

One way to study these correlations is to understand whether light-curve properties of SNe can be traced to circumstellar interactions around the SN or to the interstellar medium of the host galaxy itself.  \cite{Phillips2013} analyzed the Na I spectral lines and found that extinction is mainly due to the interstellar medium of the host galaxies and not the circumstellar material.  A related approach is to measure properties of the galaxy that are local to the SN position (\citealp{Rigault2013}, Kelsey et al. 2020 in prep.).  Still, whether local galaxy properties or global galaxy properties are more correlated with distance modulus residuals remains unclear \citep{Jones18global}.  

We introduce a new approach to study the relationship between SN light-curve properties and their host galaxies by systematically searching for galaxies that host multiple SNe Ia.  While the canonical rate of all SNe is roughly 1 SN per galaxy per 100 years, as surveys like Pan-STARRS1 (PS1) and The Dark Energy Survey Supernova Program (DES-SN) monitor a million galaxies over a 5 year time-span, the number of galaxies that host multiple SNe Ia can be significant.  In fact, \cite{Anderson2013} queried records of SN observations over 100 years and found 210 galaxies that hosted multiple SNe, though only roughly half of the SNe in these galaxies are Type Ia.  Recently, \cite{Stritzinger2010} studied 4 SNe Ia in NGC 1316 (Fornax A) and for 3 of the them 
(SN 1980N, 1981D, 2006dd) measured consistent distance moduli within 0.2 mag and having uncertainties of $0.05-0.1$ mag.  However, one of them (SN 2006mr) was fast-declining, sub-luminous and the distance modulus was 0.6 mag from the other three with a similar distance modulus uncertainty.  Similar studies have been done \citep{Gall18, Ashall18} for two SNe Ia in another galaxy (NGC 1404) in the same Fornax cluster.

While a combined historical set could potentially provide an excellent dataset to compare properties of SNe that share the same host, it is difficult to collect all the light curves of past SNe, re-calibrate them on a homogeneous system (e.g., \citealp{Scolnic2015}) and correct for selection effects.  Instead, for the present analysis we focus on the preliminary DES-SN 5-year photometrically identified SNe Ia sample, which has created one of the largest SNe Ia samples to date.  Having a well characterized telescope and survey, we are able to determine the number of galaxies that host multiple SNe.

The organization of this paper is as follows.  In section 2, we discuss the DES sample, host-galaxy association, and present the number of galaxies with multiple SNe and multiple SNe Ia.  In section 3, we compare the light-curve properties of the matched SNe Ia.  In section 4, we forecast numbers of host galaxies of multiple SNe discovered by LSST and present our conclusions.

\section{Finding Galaxies That Hosted Multiple Type Ia Supernovae}

\subsection{The DES-SN Photometric Sample and Selection}

We analyze the full, preliminary, photometric SN sample from DES-SN that was collected over 5 observing seasons spanning roughly mid-August to mid-February starting 2013.  Observations were taken with the Dark Energy Camera \citep{decam} at the Cerro Tololo Inter-American Observatory.  Details of the survey operations are given in  \cite{Kessler2015} and \cite{D'Andrea18}. The observations were taken with $griz$ passbands and in total, there are 10 fields, 8 of which are `shallow' ($r$-band $5\sigma$ visit depth of 23.4 mag) and 2 of which are `deep' ($r$-band $5\sigma$ visit depth of 24.6 mag).  We use the photometry from the DiffImg pipeline as described in \cite{Kessler2015}.  \cite{Brout18-SMP} present an improved scene-modeling photometric pipeline, and show that the DiffImg photometry is consistent to 1-2\%, which is adequate for this sibling analysis.  Most of the subtraction artifacts were rejected with a machine-learning algorithm \citep{Goldstein2015}. 

For DES-SN, detections within $1''$ of one another are grouped as a single SN candidate, and for each candidate,  PSF-fitted photometry measurements were done for all observations regardless of their signal-to-noise ratio (SNR).  We require that to be called a SN, there must be observations in two filters with SNR$>5$; this yields a total sample size of 9,289 transients.  The sample still contains a considerable number of active galactic nuclei (AGN) and image artifacts, so further vetting is needed with SN classifiers, as discussed in Section 2.3.


\subsection{Galaxy Association}

To determine which galaxies host the discovered SNe, we use co-added templates built from multiple observations.  While a relatively shallow co-add was used to build a galaxy catalog for matching SNe during DES-SN operations, in this analysis we use much deeper templates as presented in (\citealp{Wiseman2020}, hereafter W20). These templates are created for each SN from all the images taken throughout DES-SN except for the images taken within six months of the date of peak brightness of that SN. The $r$-band depth of the shallow field  templates is $r\sim25.75$ and $r\sim26.75$ mag for the deep field templates.  With the stacked templates, we associate the host galaxy following the Directional Light Radius (DLR) method \citep{Sullivan06}.  For all galaxies within $15''$ of the SNe, the shape of each galaxy is measured from SExtractor \citep{Bertin96, Holwerda} and we measure the distance between the SN position and the center of the galaxy after accounting for the galaxy shape in the direction of the SN ($d_{\rm DLR}$).  The galaxy with the smallest $d_{\rm DLR}$ is assigned as the host galaxy.

The likelihood of incorrect galaxy association is discussed using simulations with simulated galaxy catalogs in \cite{Gupta2016} and is expected to be $\sim4\%$ for $d_{\rm DLR}<4$.  This effect can also be analyzed from results of a host redshift follow-up campaign by the OzDES survey, as described in \cite{Yuan15} and \cite{Childress17}.  OzDES had cumulative redshift efficiency of $63\%$ for galaxies up to a cutoff of $r\sim24$ mag.  While the host-galaxy association followed the same procedure as described above, OzDES used
positions of host galaxies derived from 
1-mag shallower templates that were created for the DES SVA1-GOLD galaxy catalog.\footnote{Data from the DECam Science Verification period is available at: \url{https://des.ncsa.illinois.edu/releases/sva1}.}   Only $1\%$ of the host-galaxy identifications changed after using deep-stack templates in W20, which indicates that the mis-association may be lower than that found in \cite{Gupta2016}.

\begin{figure}
\centering
\includegraphics[width=0.4\textwidth]{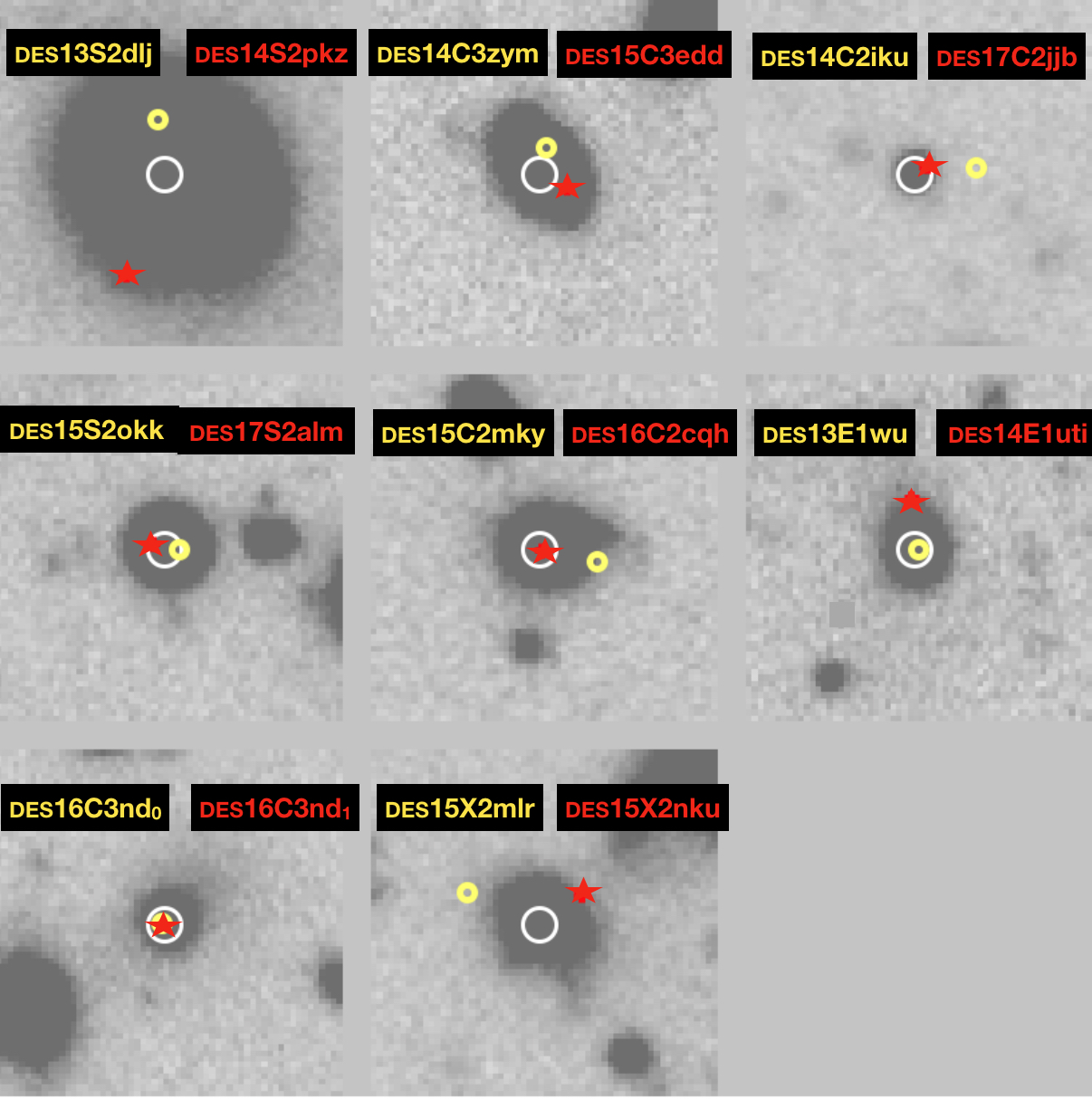}
\caption{\label{fig:mreg}Images of the host galaxies (position circled in white) of the supernova siblings with the position of the SNe marked in yellow and red. The angular scale on these plots is $16''$ on each side of each subpanel.}

\end{figure}

\subsection{Classification}

We search for galaxies that host two SNe Ia.  In total, there are 73 galaxies that host SN candidate pairs, where each SN in the pair is clearly not an AGN (classified by $>5\sigma$ non-zero flux over multiple years with positions within $1''$ of center of host galaxy) or image artifact as flagged by the difference imaging pipeline.  \color{black} Here, we use classifiers to identify SNe Ia, and assume any left-over AGN and image artifacts will not be confused with SNe Ia.  We use both the SuperNNova classifier (SNN, \citealp{Moller19}) and the PSNID classifier \citep{Sako11}.  We run the classifiers on the set of 73 transients and we do not use redshift information in the classification fits.  SNN is based on a recurrent neural network (RNN) that is trained to classify photometric light curves and returns a probability of whether a SN is type Ia or non-Ia.  We use the SNN `Vanilla' classifier and a probability threshold of 0.8. The PSNID classifier compares the SN light curves to a grid of templates that includes multiple SN types (Ia, Ibc and II) and returns a Bayesian probability based on the grid comparisons for each SN type. We use a SNe Ia probability threshold of 0.8 \citep{Sako11}.  From simulations of SNN and PSNID (Hinton et al. 2020 in prep.), the probability threshold used for each of these classifiers is in good agreement with the purity of a sample cut for that threshold.

We find good agreement between the classifiers in that they both classify the same 7 pairs of SN siblings as being two SNe Ia. Of these 14 SNe that pass the SNN threshold, 
SN DES14C2iku has a probability of 0.82, and the the SN with the next lowest probability is 0.94 (SN DES15X2mlr); these two SNe are not part of pairs that indicate disagreement in SN properties, as discussed in next section.  All of the other probabilities are above 0.985, indicating a high likelihood of being SNe Ia.  From PSNID, of the 14 SNe, the lowest probability is for SN DES16C2cqh at 0.94.  The PSNID classifier points to one additional pair (SN: DES13X3han, DES16X3eom) which is at $z=0.953$, and likely the SNR of the light-curve observations is too low for SNN to return a high likelihood classification. As one but not both of our classifiers call this a SNIa, we do not include this pair in our sample.

For PSNID, we find that including a redshift prior does not change the classifications for our sibling sample.  We also find that with the exception of one pair (DES14C2iku and DES17C2jjb), the redshifts returned from the PSNID fits are within $\Delta z$ of 0.1 from each other, which is bigger than the returned redshift uncertainties.  Furthermore, with the exception of DES14C2iku and DES13E1wu, the redshifts returned from the PSNID fits are all within $\Delta z$ of 0.1 from their host galaxy redshifts as well.


\begin{table*}
\begin{center}
\caption{\label{tab:sumtab}Summary Information about the SN candidates}
\begin{tabular}{l|l|l|l|l|l|l|l|l|l|l|} 
SN$_1$  & SN$_2$ & $z$$^{a}$& Host Pos.$^b$ & SN$_1$ Pos. & d$_{1\rm DLR}$ & SN$_2$ Pos. & d$_{2\rm DLR}$ & MJD$_1$$^{c}$ & MJD$_2$ & Mass$^{d}$\\
\hline
DES13S2dlj & DES14S2pkz & 0.228 & 40.8636   $-$01.6024 & 40.8637   $-$01.6017 & 1.145 & 40.8641   $-$01.6036 & 2.604 & 56541 & 57004 & $11.23$\\
DES14C3zym & DES15C3edd & 0.349 & 53.2948   $-$27.9576 & 53.2948   $-$27.9573 & 1.398 & 53.2945   $-$27.9578 & 1.738 & 57002 & 57286 & $10.70$\\
DES14C2iku & DES17C2jjb & 0.384 & 54.4053   $-$28.3102 & 54.4045   $-$28.3102 & 1.846 & 54.4052   $-$28.3102 & 0.232 & 56955 & 58146 & $10.31$\\
DES15S2okk & DES17S2alm & 0.506 & 41.7614   $-$01.3781 & 41.7612   $-$01.3781 & 0.618 & 41.7616   $-$01.3781 & 0.412 & 57393 & 58005 & $10.85$\\
DES15C2mky & DES16C2cqh & 0.524 & 55.1459   $-$28.6279 & 55.1451   $-$28.6281 & 2.110 & 55.1459   $-$28.6279 & 0.074 & 57348 & 57689 & $11.14$\\
DES13E1wu & DES14E1uti & 0.561 & 06.8201~   $-$42.5739 & 06.8200   $-$42.5739 & 0.185 & 06.8201   $-$42.5732 & 1.901 & 56550 & 57046 & $10.97$\\
$^e$DES16C3nd$_0$ & DES16C3nd$_1$ & 0.648 & 52.2183   $-$27.5744 & 52.2183   $-$27.5744 & 0.059 & 52.2183   $-$27.5744 & 0.059 & 57635 & 57753 & $11.14$\\
DES15X2mlr & DES15X2nku & 0.648 & 35.4094   $-$05.7659 & 35.4103   $-$05.7656 & 2.898 & 35.4088   $-$05.7656 & 2.221 & 57345 & 57363 & $9.960$\\
\hline
\end{tabular}
\end{center}
$^a$Redshift of host galaxy; the uncertainties on $z$ are $<0.001$; $^b$Positions in degrees. $^c$Date of peak brightness; uncertainties shown in Fig. 2.  $^d$Host mass such that $\log_{10}(\mathcal{M}_{\rm stellar}/\mathcal{M}_{\sun})$; uncertainties are all $\sim0.05$. $^e$SNe given same SN id because located within $1''$.
 \end{table*}

Since detections within $1''$ are assigned to the same candidate, we also search for two SN candidates within $1''$ that were assigned to a single candidate.  We use the PSNID classifier on data from each year and search for SNe Ia that appeared in multiple years but are located within $1''$.  We do not find additional candidates for SN siblings.

In the course of the survey, one event --- DES16C3nd --- was manually discovered to have two SNe within $1''$ in the course of 200 days, in the same season.  We classify both of these separately with PSNID and find both to be SNe Ia.  This 8th SNe Ia pair is included in our set.  No other distinct SNe have been discovered at the same position in the same year, but this has not been completely vetted due to difficulty in identifying these candidates.  In the Appendix, we discuss whether a discovery of 8 SN siblings is consistent with expectations from rates of SNIa per galaxy.

The 8 pairs of siblings are presented in Table~\ref{tab:sumtab}.  
The positions of the SNe relative to their host galaxies are shown in snapshots of the host galaxies from the deep-stack images in Fig. 1.  The positions and DLR values are given in Table 1.  The median $d_{\rm DLR}$ is $2.1$, within the high-accuracy range ($>97\%$) for galaxy association \citep{Gupta2016}.  We also present the host-galaxy masses in Table 1, following the prescription in \cite{Smith2020}, and similar to that done in \cite{Brout18-SYS}.  Seven of the eight hosts have mass $M > 10^{10} M_{\sun}$.

\section{Comparing Matched SNe}

\subsection{Light-curve properties and distance modulus estimates}

We fit the light curves of the 16 SNe Ia using the SALT2 model \citep{Guy2010} with the latest update from \cite{Betoule2014} as implemented in SNANA \citep{Kessler2009}.  OzDES has measured host-galaxy redshifts for all 8 galaxies, and those redshifts are used in the fits.  The light-curve fits return: an overall amplitude parameter $x_0$, which can be converted to a brightness $m_{\rm B}$; $x_1$, the light-curve stretch; and $c$, the light-curve color.  The light-curve fits are shown in Fig. 2 and a comparison of the fitted parameters for each pair siblings is shown in Fig. 3.   

To convert the fitted parameters to a distance modulus measurements, we follow the Tripp estimate \citep{Tripp98},
\begin{equation}
    \mu = m_{\rm B} +\alpha x_1 - \beta c - M ,
\end{equation}
where $\alpha$ and $\beta$ are the correlation coefficients of luminosity with $x_1$ and $c$ respectively, and $M$ is the absolute magnitude of SNe Ia.  From \cite{Brout18-SYS}, we use $\alpha=0.14$ and $\beta=3.1$ for this analysis.  We account for Milky-Way extinction in our light-curve fits with values from \cite{Schlafly11}.  We do not apply additional bias corrections from the BBC method \citep{BBC}.  The $\mu$ comparisons are also shown in Fig. 3.  Since we are investigating the intrinsic scatter of SNe Ia, we do not include an additional intrinsic scatter term in the uncertainty on $\mu$.  However, we do include the default SALT2 model error in our distance modulus uncertainties as the model error is included as part of the measurement uncertainties.

\begin{figure*}
\centering
\includegraphics[width=1.0\textwidth]{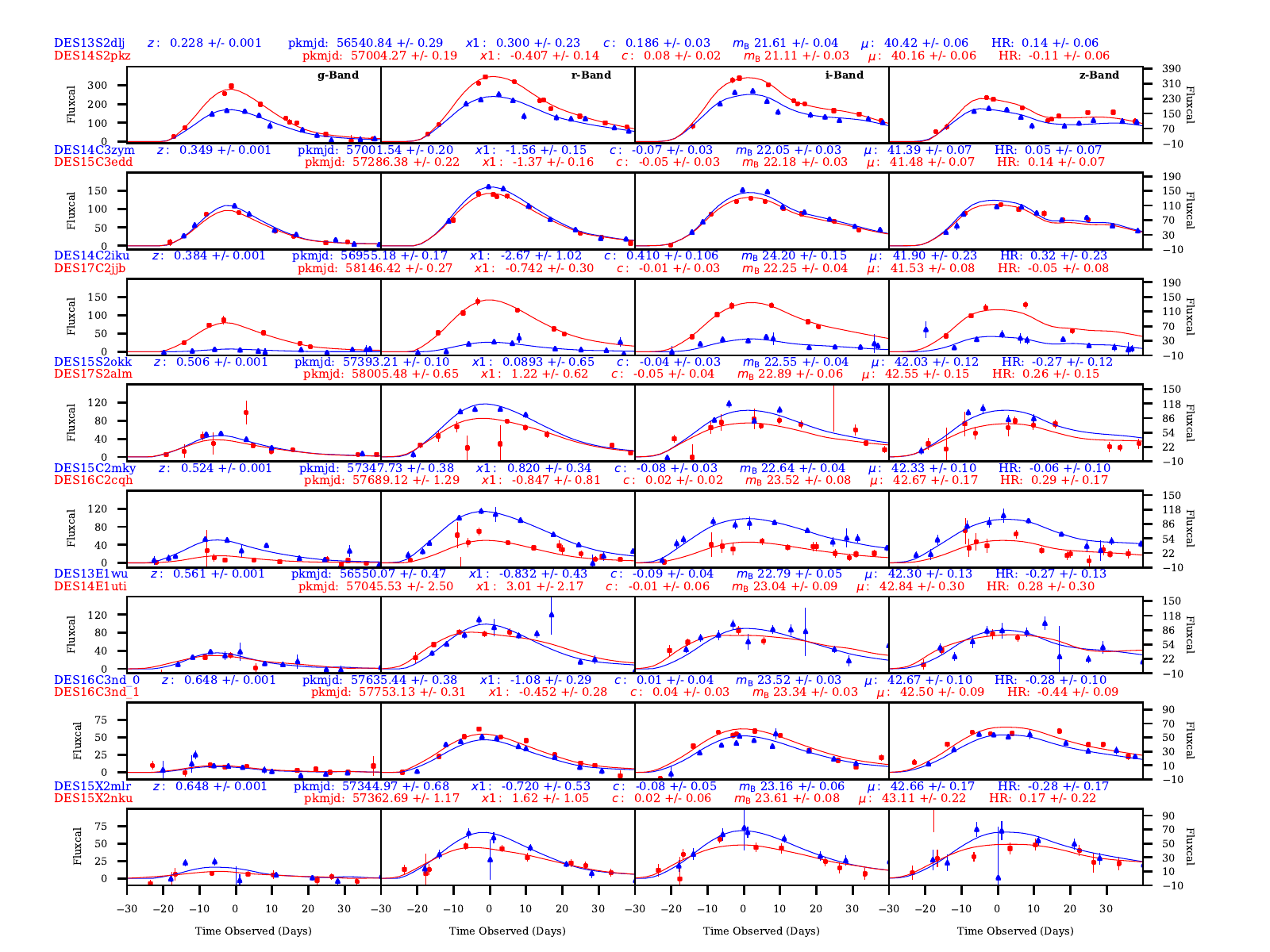}
\caption{\label{fig:sib}Light curves of each pair of SNe Ia siblings.  The SN name, redshift ($z$), date of peak brightness (peak mjd), light-curve stretch ($x_1$), light-curve color ($c$), relative brightness ($m_{\rm B}$) and distance modulus ($\mu$), and Hubble residual to a fiducial cosmology (HR) are shown for each.  The uncertainties on $\mu$ and HR are shown assuming no intrinisc scatter.} 
\end{figure*}

\begin{figure}
\centering
\includegraphics[width=0.5\textwidth]{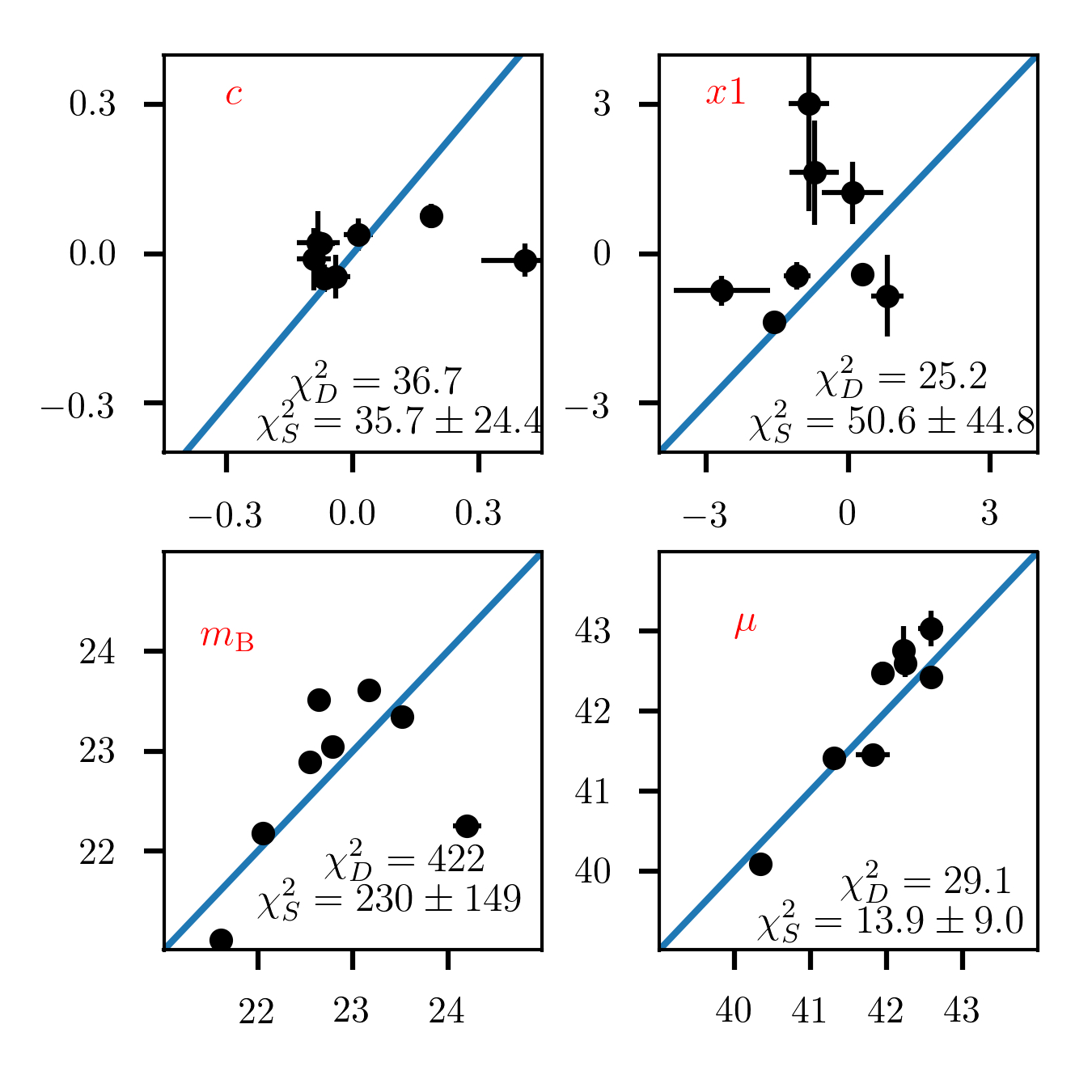}
\caption{\label{fig:lightcurves}A comparison of the light-curve parameters ($c$, $x_1$ and $m_{\textrm{B}}$) and the distance modulus $\mu$ of the matched SNe.  The SN chosen for the $x$ versus $y$ values is arbitrary and can be flipped.  The $\chi^2_D$ of the data points to the $y=x$ line is given. The mean $\chi^2_S$ predicted from an ensemble of simulated samples of the same size and with a full intrinsic scatter model is also given, as well as the width of that distribution, given as $1\sigma$.}
\end{figure}
\subsection{Assessing consistency with simulations}

 To provide context to the comparison of the light-curve properties of different SNe from the same hosts, simulations are needed.  We follow \cite{Kessler18} for DES-SN simulations, with two modifications.  First, we use the 5-year observing history from DES-SN rather than the 3-year history.  Second, we do not include a spectroscopic SN efficiency in our model as we are analyzing the full photometric sample.  In the analysis of the simulated sample, we apply the same light-curve quality requirements (cuts) as described in \cite{Brout18-SYS} to the simulations as we do to the data, with the exception of light-curve $c$ range, where we loosened this cut for the data to include one more SN.  While we fortuitously have host-galaxy redshifts for our entire sibling sample, we did not require a spectroscopic host galaxy redshift measurement, and therefore we do not include a redshift efficiency model in our simulations.  
 
 The intrinsic scatter model used in our simulations is from \cite{Guy2010}, as adapted in \cite{Kessler13}.   We would like to simulate a single sample with uncorrelated intrinsic scatter for most SNe, but correlated to a varying amount amongst SNe with the same hosts.  Since this is difficult to implement, we instead simulate three independent samples where we scale the magnitude of the entire intrinsic scatter from 1 (here called SIM-1) to $1/2$ (SIM-1/2), which is half the magnitude of intrinsic scatter, to 0 (SIM-0), which is no intrinsic scatter.  This method assumes that the amount of the intrinsic scatter not accounted for is 100\% correlated between SNIa with the same host.  While we don't explicitly simulate the correlated component of the intrinsic scatter, this should have a negligible impact on the the analysis.

We create a simulated sample with 400,000 DES SNe and select 8 random pairs of SNe, where pairs are defined as being within 0.05 in redshift, where we use the redshifts of the 8 host galaxies given in Table 1 and 0.05 is an arbitrary bin size that ensures similar noise properties for light curves of SNe Ia with $z>0.2$.  When comparing $m_{\textrm{B}}$ or $\mu$, we subtract the cosmological dependence of the SNe due to different redshifts.  For each pair, and for each parameter, we define the $\chi^2$ with $\chi^2_D$ for data and $(\chi^2_S)$ for simulations where
\begin{equation}
    \chi^2=\sum_i^8 ({O1-O2})^2/({\sigma_{O1}^2+\sigma_{O2}^2})
\end{equation}
where O1 and O2 are the observables for each pair ($x_1,c,m_{\rm B},\mu$) and $\sigma_{O1}$ and $\sigma_{O2}$ are the uncertainties on those observables.  This process of pulling 8 pairs is repeated 1000 times.

We give the $\chi^2_D$ and $\chi^2_S$ for each $m_{\textrm{B}}$, $x_1$, $c$ and $\mu$ in Fig. 3.  We find that for $c$, $x_1$ and $m_{\textrm{B}}$, the reduced $\chi^2_S$ are predicted from simulations to be $>>1$ as the parameters are drawn from distributions of width significantly larger than the uncertainties. We find that only for $x_1$ is $\chi^2_{D}$ smaller than $\chi^2_{S}$.  This smaller value implies that SNe Ia in the same hosts have similar $x_1$ values.  We also find that $\chi^2_{D}$ in $\mu$ is higher than $\chi^2_{D}$ for $m_{\textrm{B}}$, which shows the impact of the light-curve standardization and that these pairs of SNe likely do have the same host galaxies.

\color{black}
The distributions of $\chi^2_{S}$ from each simulated sample of 8 siblings are shown in Fig. 4, along with the $\chi^2_D$.  For both data and sims, the $\chi^2$ is calculated using measurement uncertainties alone and not including any additional intrinsic scatter uncertainty.  We find $\chi^2_D$ in $\mu$ is $29.1$, which we can see is inconsistent with SIM-0, but is consistent with SIM-1.  We can place a constraint from this comparison by converting the probability that $\chi^2_S$ is above $\chi^2_D$ by computing the inverse cumulative normal probability deviation for a given cumulative probability \citep{PhysRevD.98.030001}.  We use SIM-1/2, and as can be seen in Fig. 4, the $\chi^2_D$ of the data is in the $<1\%$ high end of the SIM-1/2 distribution.  From this, we find that at $2.8\sigma$, we can rule out global host-galaxy properties causing more than 1/2 of the total intrinsic scatter of SNe Ia Hubble residuals.  

The high $\chi^2_D$ from the data for $
\mu$ is driven by two of the pairs, 
(DES15S2okk, DES17S2alm) and (DES13S2dlj, DES14S2pkz) with $\chi^2_{D}=7.4, 8.5$ respectively.  The 4 SNe from these pairs do not appear unusual in any way.  The other matches all have $\chi^2< 3.1$. One pair with a low $\chi^2_D$ of 2.3 is (DES14C2iku, DES17C2jjb), one of the two SNe in this pair (DES14C2iku) has a $c$ value of $0.41\pm0.11$, which would be cut in a typical cosmology analysis (e.g. \citealp{Scolnic18}).  We include it here as it passed classification, has a sibling, and has large uncertainties; removing it would not change our conclusions.

\begin{figure}
\centering
\includegraphics[width=0.5\textwidth]{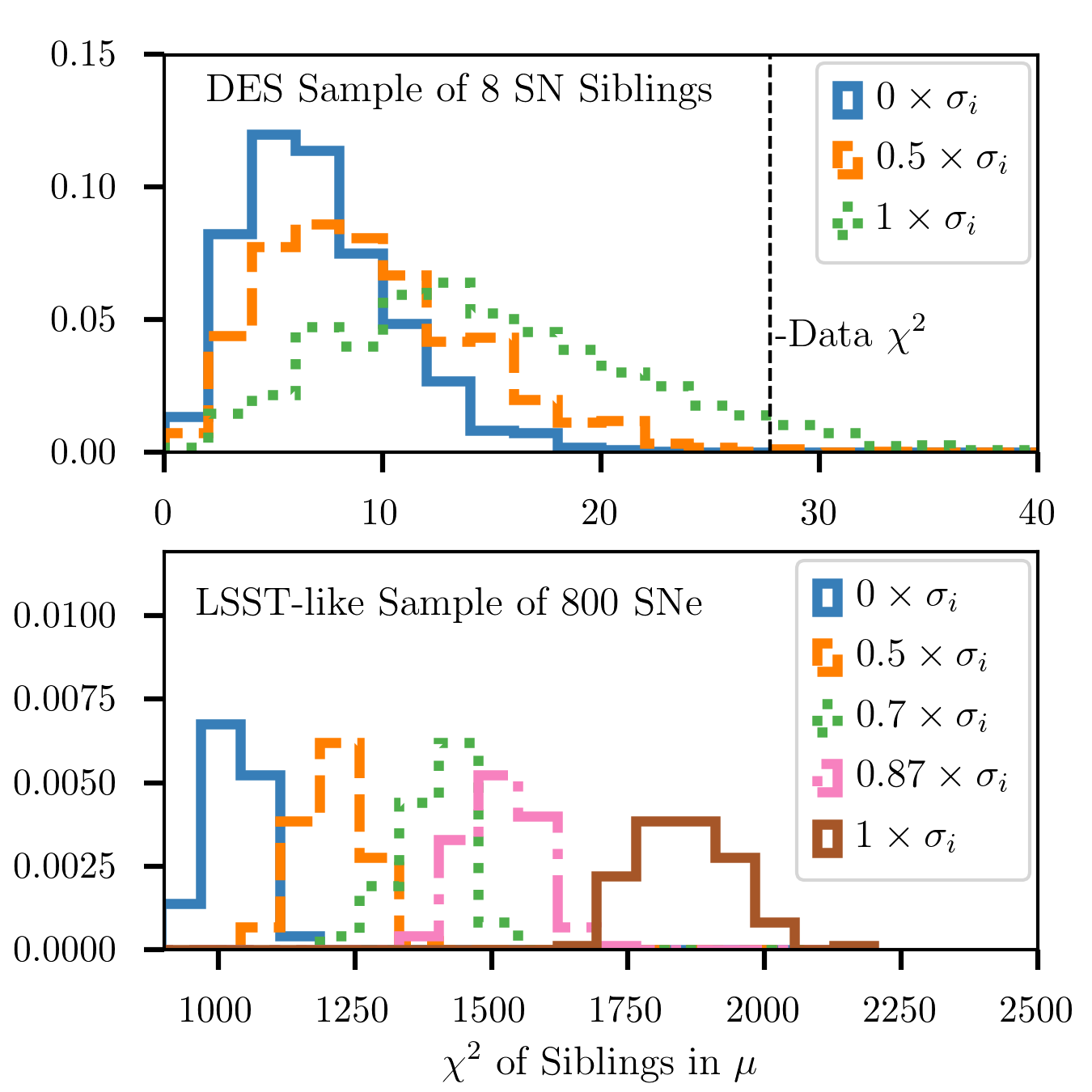}
\caption{\label{fig:distribution}(Top) From evaluations of the distance modulus $\mu$, the predicted $\chi^2_S$ distribution from simulations of 8 DES SN siblings and the $\chi^2_D$ from the real data set - as presented in Fig. 3.  The three histograms show distributions of $\chi^2_S$ of 1,000 samples of 8 siblings from simulations with no intrinsic scatter (blue), half intrinsic scatter (red) and full intrinsic scatter (green).  (Bottom) Again for $\mu$, the predicted $\chi^2_S$ distribution from simulations of 800 pairs of LSST SN siblings.  Similar to above, the multiple histograms are created for simulations with different scales of the intrinsic scatter.}
\end{figure}

\subsection{The Likelihood of Lensed Supernovae}

Two of the supernova sibling pairs have members that exploded within 100 days of each other, which raises the possibility that we may have discovered lensed SNe.  The first pair is DES15X2mlr and DES15X2nku, where the dates of peak brightness differ by 20 days and the SNe are on opposite sides of the galaxy.  The DES-SN deep stack shows ring-like structure around the galaxy.  However, the Hubble residuals for both SNe do not show evidence of magnification ($\Delta \mu=-0.28\pm0.20$ mag, $\Delta \mu=0.17\pm0.24$ mag) which decreases the likelihood that the SNe are lensed.  Additionally, if we use PSNID to perform a photo-$z$ fit to the SN light curves itself \citep{Sako11}, we recover best-fit redshifts ($z=0.633, 0.555$ with uncertainties $\sigma_z\sim0.05$) which are consistent with the host-galaxy redshift $z=0.648$.  This agreement is not expected for lensed SNe Ia. Additionally, the relatively low mass ($<10^{10}$ $M_{\odot}$) of the host galaxy makes it unlikely that there is a lensed SN with source-lens separation of this size ($\sim3''$). 

The second pair was identified as the same SN --- DES16C3nd --- given the locations of the two SNe are within $1''$.  Unlike the previous pair, the Hubble residuals are more negative ($\Delta \mu=-0.28 \pm 0.14$, $\Delta \mu= -0.44 \pm 0.14$), which shows $\sim2-3\sigma$ hints of magnification, when assuming 0.1 mag of intrinsic scatter. However, OzDES acquired a spectroscopic redshift of the SN itself with AAT of the second SN (on Dec. 30 2016) and at $z=0.65$ it is in good agreement with the redshift of the host galaxy at $z=0.6483$, reducing the chances that the SN is lensed.  While it appears to be unlikely that two SNe would  appear within 100 days in the same galaxy within $1''$, further follow-up would be needed to help understand if this is purely coincidental.  Still, a separation near $0''$ indicates that there is no lensing taking place.

\section{Discussion and Conclusion}

Here we find 8 galaxies that host multiple Type Ia SNe and make the first quantitative consistency test of SNe Ia light-curve properties for a sample of SNe Ia that share the same host galaxy.  Overall, we find that at most $1/2$ of the intrinsic scatter of SNe Ia Hubble residuals can be contributed from the parent galaxy.  Only for the light-curve property $x_1$ do we see weak evidence that the stretch is drawn from a more narrow population than for the full sample of SNe Ia.

The main result from this analysis does not contradict the various correlations between SNe Ia luminosity and host-galaxy properties found in the literature.  We check the recent analysis of Pantheon \citep{Scolnic18} combined with Foundation \citep{Foley2018} done in \cite{Kenworthy19}, and find that the recovered mass step of $\gamma=0.052$ reduces the intrinsic scatter of the sample compared to one without a mass step from $0.105$ to $0.101$, or $\sim4\%$ of the total intrinsic scatter.  This reduction is less than $1/2$ of the intrinsic scatter constrained from our analysis, and we would need more SN siblings to measure the reductions of intrinsic scatter due to additional host standardization parameters.

There are two main systematic uncertainties with our approach.  The first is incorrect galaxy association.  Since this is accurate to the $97\%$ level and the Hubble residuals are all low in magnitude, the likelihood of mis-association is probably low, however we cannot rule this out.  As our sample is small, a single mis-association could have a strong impact on our conclusions.  A second systematic is that the SNe identified are not separate SNe but actually a single lensed SN.  This is discussed above, and if it was the case, it should reduce the total $\chi^2_D$ as given in Fig. 4 as the properties of the multi-imaged SNe Ia should be more similar than two random SNe Ia \citep{Goobar17}.  Still, a follow-up observing program that utilizes a telescope with high spatial resolution would be useful to better understand if there are lensing effects.  Further systematic uncertainties reside in the treatment of the simulations; for example, simulating $x1$ and $c$ distributions that correspond to SNe in high-mass galaxies could improve analysis about agreement of these properties in same-host galaxies.

While the constraint on the contribution to intrinsic variance from having the same parent galaxy is not very strong for this small sample, a much more significant statement can be made with LSST.  Following the simulations as described in \cite{Kessler19} and using selection requirements detailed in \cite{Mandelbaum18}, we calculate that there will be roughly 300,000 SNe with similar SNR cuts as discussed above over a similar redshift range.  As the rate of SNe per galaxy will be the same for DES-SN and LSST, we scale the number of SNe Ia siblings by $100\times$.  This scaling results in $\sim800$ pairs of sibling SNe Ia, roughly equal to our largest SNe Ia samples \citep{Scolnic18}.  In Fig. 4, we show the ability to constrain the intrinsic scatter of SNe Ia Hubble residuals with LSST.  This analysis should be able to place constraints to a quarter of the total variance.  Assuming that we will have spectroscopic redshifts for all our host galaxies, additional spectroscopic follow-up programs for the second siblings  would be useful for better understanding this sample.  

Although we  evaluate only the correlation of global properties with this approach, it will be likely that that a fraction ($\sim10\%$) of the pairs will occur in the same galaxy at similar positions to within $2''$, and similar analyses attempting to study both global and local properties can be achieved.  In our sample, for the two pairs of SN siblings with the most different distance modulus values, one pair has both SNe located near the center of the galaxy, and the other pair has the SNe located on the opposite sides of the galaxy. It is also interesting to note that in our DES-SN sample of siblings, all but one of the host galaxies has a high mass ($>10$), as can be seen in Table 1.  It is possible that the higher Hubble residual scatter is due to these SNe being in more massive galaxies.

Finally, we remark on an analysis that could be done with past surveys.  As mentioned in the introduction, there are some well known systems, such as Fornax, that have hosted multiple SNe Ia.  We searched the PS1 SN database \citep{Jones18}, and find no candidates for supernova siblings.  The total number of SNe in the PS1 sample relative to DES-SN is $3\times$ fewer, and therefore should expect on the order of 2-3, so 0 pairs of siblings is consistent with Poisson noise.  Similarly, we find no siblings in the full Pantheon set of 1048 SNe.  Recently, SN 2017cbv appeared in the same galaxy as SN 2013aa, and this pair will be used in the SH0ES analysis for measuring the Hubble constant \citep{Riess16}.  Understanding the limited correlation between SNe in the same host galaxy will be important for SH0ES to properly propagate the combined uncertainty from this pair of siblings. 

In conclusion, finding SN siblings is an exciting avenue for improving systematics in cosmology studies with SNe Ia, and should be particularly promising in the LSST era.
\appendix

\section{Agreement with Expected Rates}

To check our sample size, we use our discovery rate of SNe Ia siblings to derive a predicted rate of SNe Ia per galaxy and compare to values from the literature.  \cite{Li2011} find from the volume-limited LOSS survey that the rate per galaxy is $0.54\pm0.12$ SNe Ia per $10^{10}$ solar masses per century.  While the masses of host galaxies has not been evaluated for the full DES-SN photometric sample, we use the methodology described in \cite{Smith2020} for the 1,934 transients in W20 identified as likely SNe Ia for which a spectroscopic host-galaxy redshift has been acquired.  The mean mass of the W20 sample is $\log_{10}(\mathcal{M}_{\rm stellar}/\mathcal{M}_{\sun})= 10.35$.  Systematic uncertainties in this estimate of the mean mass for the full sample are due to preferential selection of brighter, and more massive galaxies for the AAT sample, but also higher identification of redshifts for more star-forming, and therefore typically less massive galaxies.  While these two effects likely don't cancel, we still use this estimate for a rough comparison. The \cite{Li2011} rate for our sample is therefore $10^{10.35}/10^{10}\times(0.54\pm0.0005)=1.21\pm0.27$ SN per galaxy per century, or $0.012\pm0.0027$ SN per galaxy per year.  This rate is given for $z=0$, and as discussed in \cite{Perrett12}, we must divide this rate by $(1+z)$ where $z$ is the mean redshift of discovered SNe Ia in the sample. DES-SN discovers SNe Ia with typical $z$ of $z\sim0.5$, from which we calculate a rate of $0.008\pm0.002$ SN per galaxy per year.

The discovered rate of SNe Ia siblings from our sample can be expressed as
\begin{equation}
\begin{split}
    \textrm{Rate}=\textrm{Pairs/TotalHosts/Surveytime/Efficiency} .\\
\end{split}
\end{equation}
For our sample, the number of pairs is 8 and the survey time is 2.3 yr, as calculated by the sum of the 5 season lengths of DES.  We use SNN classifier on the DES-SN photometric sample with a probability threshold of 0.8 and find the number of total hosts to be 3,227.  The efficiency is much more difficult to calculate. From simulations described in \cite{Kessler18}, the SN discovery efficiency is $0.15\pm0.01$. From Eq. A1, the SNIa rate is $\sim0.007$ SN per galaxy per year.

This rate is in good agreement with the prediction from \cite{Li2011} of $0.008\pm0.002$ SN per galaxy per year.  The main limitation of the calculation is how to properly account for selection effects.  Here we took the discovery efficiency of SNe given a redshift distribution from $0.1<z<1.2$, but this distribution is not the same as the redshift distribution of our 8 siblings.   Still, while a full rates analysis is beyond the scope of this paper, this simple calculation shows that the size of the sample found is within expectations and can be used for further analysis.

\acknowledgments

DS is supported by DOE grant DE-SC0010007, the David and Lucile Packard Foundation.  DS and RK are all suported in part by NASA under Contract No. NNG17PX03C issued through the WFIRST Science Investigation Teams Programme.  DB and MS were supported by DOE grant DE-FOA-0001358 and NSF grant AST-1517742.  L.G. was funded by the European Union's Horizon 2020 research and innovation programme under the Marie Sk\l{}odowska-Curie grant agreement No. 839090. This research used resources of the National Energy Research Scientific Computing Center (NERSC), a DOE Office of Science User Facility supported by the Office of Science of the U.S. Department of Energy under Contract No. DE-AC02-05CH11231. We are grateful for the support of the University of Chicago Research Computing Center for assistance with the calculations carried out in this work.  RF is supported in part by NSF grants AST-1518052 and AST-1815935, the Gordon \& Betty Moore Foundation, the Heising-Simons Foundation, and by fellowships from the David and Lucile Packard Foundation, as well as the NASA contract above.

Funding for the DES Projects has been provided by the U.S. Department of Energy, the U.S. National Science Foundation, the Ministry of Science and Education of Spain, 
the Science and Technology Facilities Council of the United Kingdom, the Higher Education Funding Council for England, the National Center for Supercomputing 
Applications at the University of Illinois at Urbana-Champaign, the Kavli Institute of Cosmological Physics at the University of Chicago, 
the Center for Cosmology and Astro-Particle Physics at the Ohio State University,
the Mitchell Institute for Fundamental Physics and Astronomy at Texas A\&M University, Financiadora de Estudos e Projetos, 
Funda{\c c}{\~a}o Carlos Chagas Filho de Amparo {\`a} Pesquisa do Estado do Rio de Janeiro, Conselho Nacional de Desenvolvimento Cient{\'i}fico e Tecnol{\'o}gico and 
the Minist{\'e}rio da Ci{\^e}ncia, Tecnologia e Inova{\c c}{\~a}o, the Deutsche Forschungsgemeinschaft and the Collaborating Institutions in the Dark Energy Survey. 

The Collaborating Institutions are Argonne National Laboratory, the University of California at Santa Cruz, the University of Cambridge, Centro de Investigaciones Energ{\'e}ticas, 
Medioambientales y Tecnol{\'o}gicas-Madrid, the University of Chicago, University College London, the DES-Brazil Consortium, the University of Edinburgh, 
the Eidgen{\"o}ssische Technische Hochschule (ETH) Z{\"u}rich, 
Fermi National Accelerator Laboratory, the University of Illinois at Urbana-Champaign, the Institut de Ci{\`e}ncies de l'Espai (IEEC/CSIC), 
the Institut de F{\'i}sica d'Altes Energies, Lawrence Berkeley National Laboratory, the Ludwig-Maximilians Universit{\"a}t M{\"u}nchen and the associated Excellence Cluster Universe, 
the University of Michigan, the National Optical Astronomy Observatory, the University of Nottingham, The Ohio State University, the University of Pennsylvania, the University of Portsmouth, 
SLAC National Accelerator Laboratory, Stanford University, the University of Sussex, Texas A\&M University, and the OzDES Membership Consortium.

Based in part on observations at Cerro Tololo Inter-American Observatory, National Optical Astronomy Observatory, which is operated by the Association of 
Universities for Research in Astronomy (AURA) under a cooperative agreement with the National Science Foundation.

The DES data management system is supported by the National Science Foundation under Grant Numbers AST-1138766 and AST-1536171.
The DES participants from Spanish institutions are partially supported by MINECO under grants AYA2015-71825, ESP2015-66861, FPA2015-68048, SEV-2016-0588, SEV-2016-0597, and MDM-2015-0509, 
some of which include ERDF funds from the European Union. IFAE is partially funded by the CERCA program of the Generalitat de Catalunya.
Research leading to these results has received funding from the European Research
Council under the European Union's Seventh Framework Program (FP7/2007-2013) including ERC grant agreements 240672, 291329, and 306478.
We  acknowledge support from the Australian Research Council Centre of Excellence for All-sky Astrophysics (CAASTRO), through project number CE110001020, and the Brazilian Instituto Nacional de Ci\^encia
e Tecnologia (INCT) e-Universe (CNPq grant 465376/2014-2).  We acknowledge support from EU/FP7-ERC grant number 615929.

This manuscript has been authored by Fermi Research Alliance, LLC under Contract No. DE-AC02-07CH11359 with the U.S. Department of Energy, Office of Science, Office of High Energy Physics. 

Based in part on data acquired through the Australian Astronomical Observatory, under program A/2013B/012. We acknowledge the traditional owners of the land on which the AAT stands, the Gamilaraay people, and pay our respects to elders past and present.

\bibliographystyle{mn2e}
\bibliography{sample}

\end{document}